\documentclass{article}

\usepackage{PRIMEarxiv}
\usepackage{algpseudocode}
\usepackage{algorithm, algorithmicx}
\usepackage{amsmath}
\usepackage{caption}
\usepackage[utf8]{inputenc} 
\usepackage[T1]{fontenc}    
\usepackage{float}
\usepackage{biblatex} 
\usepackage{hyperref}       
\usepackage{url}            
\usepackage{booktabs}       
\usepackage{amsfonts}       
\usepackage{nicefrac}       
\usepackage{microtype}      
\usepackage{lipsum}
\usepackage{fancyhdr}       
\usepackage{graphicx}       
\graphicspath{{media/}}     
  
\title{Accelerating a Triton Fused Kernel for W4A16 Quantized Inference with SplitK work decomposition
}

\author{
  Adnan Hoque \\
  IBM T.J. Watson Research Center \\
  Yorktown Heights, NY, United States\\
  \texttt{adnan.hoque1@ibm.com} \\
   \And
  Less Wright \\
  Meta  AI \\
  \texttt{less@meta.com} \\
   \And
  Chih-Chieh Yang \\
  IBM T.J. Watson Research Center \\
  Yorktown Heights, NY, United States\\
  \texttt{chih.chieh.yang@ibm.com} \\
   \And
  Mudhakar Srivatsa \\
  IBM T.J. Watson Research Center \\
  Yorktown Heights, NY, United States\\
  \texttt{msrivats@us.ibm.com} \\
   \And
  Raghu Ganti \\
  IBM T.J. Watson Research Center \\
  Yorktown Heights, NY, United States\\
  \texttt{rganti@us.ibm.com} \\
   \And
}

\begin{document}
\maketitle

\begin{abstract}
We propose an implementation of an efficient fused matrix multiplication kernel for W4A16 quantized inference, where we perform dequantization and GEMM in a fused kernel using a SplitK work decomposition. 
Our implementation shows improvement for the type of skinny matrix-matrix multiplications found in foundation model inference workloads. In particular, this paper surveys the type of matrix multiplication between a skinny activation matrix and a square weight matrix. Our results show an average of 65\% speed improvement on A100, and an average of 124\% speed improvement on H100 (with a peak of 295\%)  for a range of matrix dimensions including those found in a llama-style model, where m < n = k. 
\end{abstract}


\section{Introduction}

There has been a large amount of research performed on designing efficient General Matrix Multiplication (GEMM) kernels \cite{noauthor_cublas_nodate} \cite{thakkar_cutlass_2023}. However, there are far fewer publications on how to design high-performance kernels for memory bound computations, such as those commonly found in inference.  For inference, matrix multiplications are usually memory bound because of the problem size (m, n and k) creating skinny matmuls where the small dimension is the batch size (i.e. 1-16), along with GPU compute and memory throughput limitations. The implementation we propose shows promising results for this case. Rather than the traditional Data Parallel (DP) decomposition, we leverage a SplitK work decomposition with atomic reductions, and integrate it with dequantization to provide a one step fused dequant and matrix multiply kernel. We implement our kernel in Triton and provide the source code on GitHub \cite{noauthor_foundation-model-stacktriton_nodate}. We chose to implement our kernel in Triton, as it provides an easy-to-use interface and also cross-hardware compatibility.  Thus, current and future iterations of this kernel can cater to the growing diversity of software/hardware stacks found in workloads across the industry.

\section{Method}

\begin{figure}[ht]
    \centering
    \begin{minipage}[b]{0.48\textwidth}
        \includegraphics[width=\textwidth]{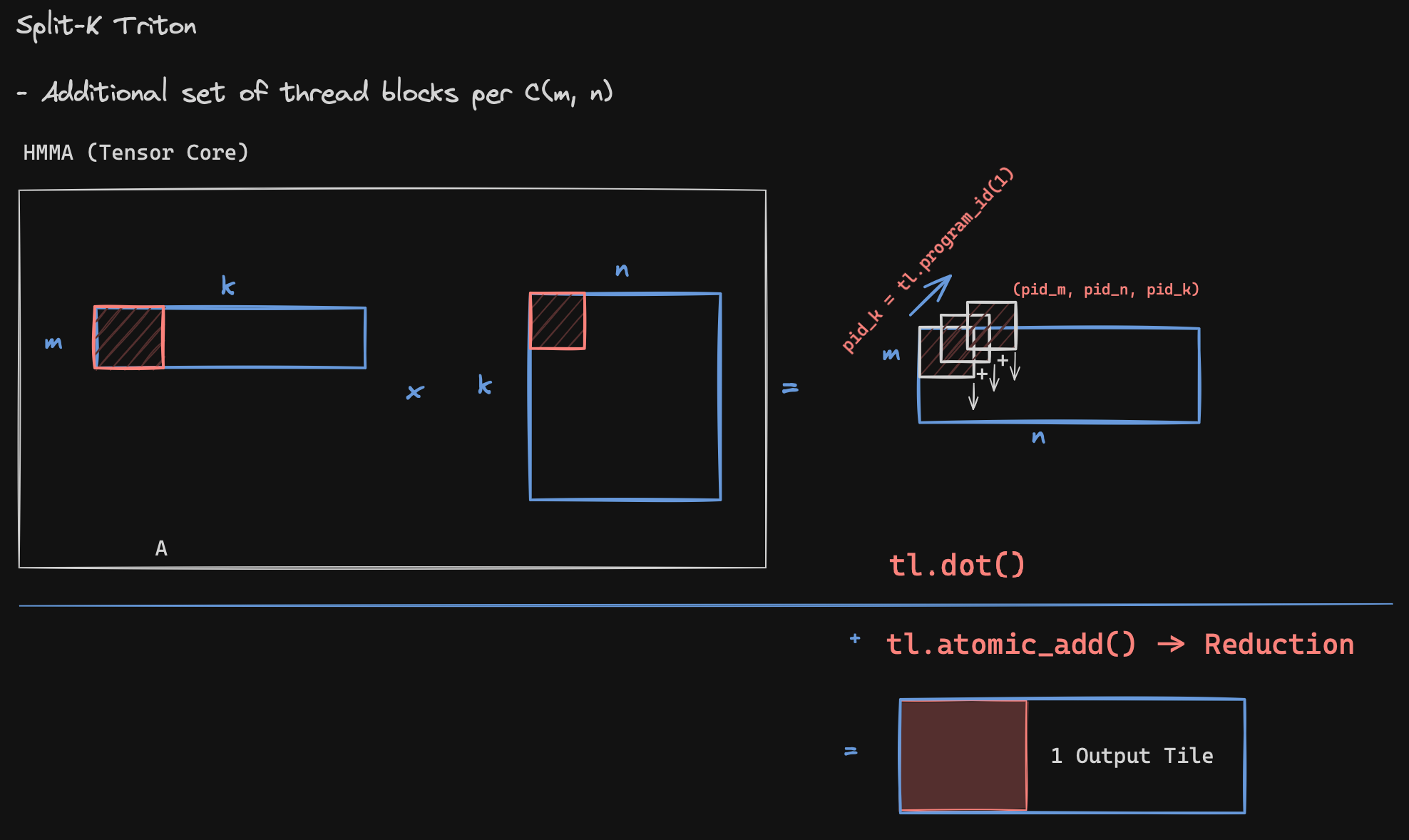}
        \caption{SplitK Thread Block Level}
        \label{fig:split_k_drawing}
    \end{minipage}
    \hfill
    \begin{minipage}[b]{0.48\textwidth}
        \includegraphics[width=\textwidth]{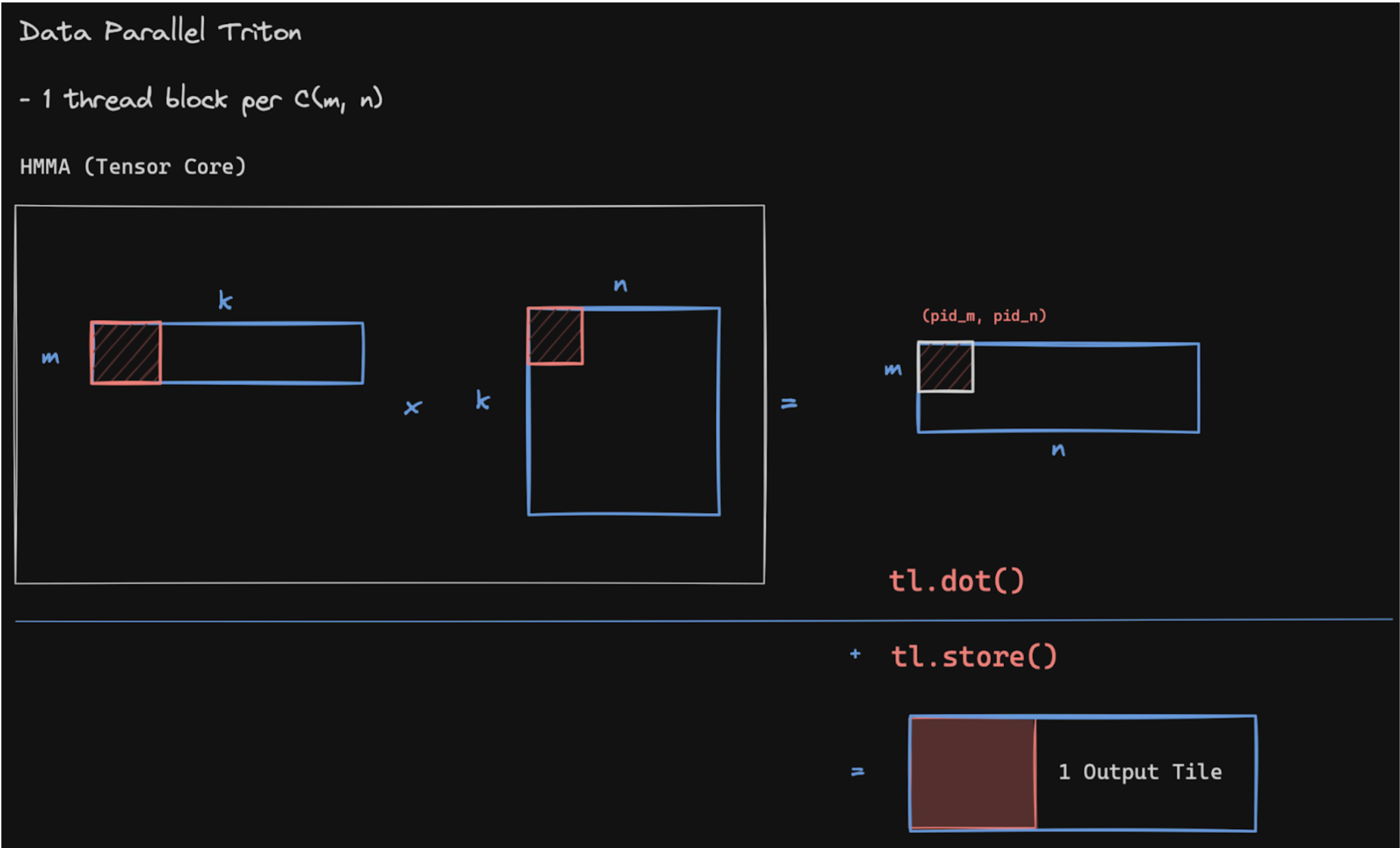}
        \caption{Data Parallel Thread Block Level}
        \label{fig:data_parallel_drawing}
    \end{minipage}
\end{figure}

Our kernel will receive as input an FP16 activation matrix, a packed (quantized) int32 weight matrix, and the scale and zero parameters that will be used to dequantize the weight matrix. The weight matrix is dequantized, scaled and shifted using bitwise operations. Our implementation is tailored to respect GPTQ-style int4 quantization, but the method is relatively general purpose and can be extended to other methods of n-bit quantization. This is then integrated with the optimized SplitK GEMM, an algorithm that first appeared in the CUTLASS library \cite{noauthor_cutlassmediadocsefficient_gemmmd_nodate}. Our kernel is inspired by the Triton FP16 SplitK implementation \cite{noauthor_tritonpythontritonopsmatmulpy_nodate}. SplitK launches additional thread blocks along the k dimension to calculate partial sums. The partial results from each thread block are then summed using an atomic reduction \ref{fig:split_k_drawing}. Thus, our implementation is a fused kernel that performs dequantization, GEMM and atomic reduction in a single kernel launch with notable performance improvements.

\subsection{Atomics - Enabling the SplitK strategy}

SplitK effectively decomposes the work into finer grained partitions as compared to traditional data parallel block style (tiling), and thus allows for more evenly balanced resource usage of the SMs.  On an A100 with our kernel, this results in waves per sm increasing by 61\% versus the standard data parallel block tiling.  This finer grained decomposition is enabled by the use of atomic operations. In our case, we leverage the atomic add function so that as thread blocks complete their inner accumulation step, which represents a partial portion of the final result, they can safely update their partial results directly to the C output memory.  This same output memory will also be updated by other thread blocks working on subsets that will contribute to the final aggregated multiply-accumulate output results for the given C output tile.  Thus, ensuring that thread blocks both update the latest results, and have exclusive access while writing their result, is paramount and achieved via atomic adds. Note that there is a tension between the improvements from finer grained SM work distribution, and the overhead of thread blocks contending for exclusive write access to the same output buffer.  This effect was seen on an A100 where increasing the SplitK parameter from 4 to 16, resulted in a steady degradation of performance as the matrix sizes increased, presumably due to greater wait times per thread block to get exclusive write access to the same memory output buffer.

\subsection{Wave Quantization effects - SplitK improves as GPU SM count improves} 

With our SplitK kernel, we saw proportionate performance gains moving from an A100 to H100 relative to the data parallel block approach.  As newer GPU's continue to have both more SM's and each SM at the same time becomes more capable, standard block tiling GEMM's will find steadily greater difficulty in avoiding the effects of wave quantization inefficiency and finding efficient work decomposition tiling sizes. To clarify, wave quantization inefficiency refers to the effect where the final wave of a tiled decomposition may only have a small subset of SM's active, with the remaining SM's idling.   This effect is usually mitigated by having oversubscription, where larger amounts of tile production work is assigned to idle SM's.  However, newer GPU's have more SM's, which means fewer waves needed to complete the desired tile ouputs/work. In addition, newer GPU's  have more powerful SM's, which require larger blocking tile sizes, which also leads to fewer waves needed.  
Execution schedules with fewer waves are much more prone to wave quantization inefficiency, as the likelihood of total output tiles being oversubscribed and scheduled as an even multiple of SM's rapidly diminishes, thus rendering the final wave likely to use only a small portion of GPU resources. This effect was indirectly shown when moving our fused kernel from A100 to H100 GPUs.  The H100 has 33\% greater SMs as well as more capable SM's.  Intuitively this should result in greater wave quantization inefficiency for data parallel, while SplitK can effectively leverage the SM improvements by simply increasing the SplitK hyperparam (from 4 on A100 to 8 on H100). As a result, compared to the data parallel approach, the average gain with our SplitK kernel on H100 nearly doubled with 124\% average performance gain, versus the A100 average gain of 65\%.

\subsection{Algorithm}

\begin{algorithm}[H]
    \caption{SplitK GEMM in Triton}
    \label{alg:matmul_split_k}
    \begin{algorithmic}[1] 
    \State Initialization of block and thread indices
    \State $pid \gets \text{tl.program\_id}(0)$
    \State $pid\_k \gets \text{tl.program\_id}(1)$
    \State $num\_pid\_k \gets \text{tl.cdiv}(k, block\_k \times split\_k)$
    \State Compute thread tile offsets
    \State $offs\_m, offs\_n, offs\_k \gets \text{ComputeOffsets}(pid, pid\_k, block\_m, block\_n, block\_k)$
    \State Initialize accumulators
    \State $acc \gets \text{tl.zeros}((block\_m, block\_n), \text{dtype}=\text{tl.float32})$
    \State Main computation loop
    \For{$k \gets 0$ \textbf{to} $num\_pid\_k$}
        \State Load input tiles and apply quantization
        \State $a, b \gets \text{LoadAndDeQuantize}(a\_ptr, b\_ptr, offs\_m, offs\_n, offs\_k)$
        \State Perform matrix multiplication
        \State $acc \gets acc + \text{tl.dot}(a, b)$
        \State Update pointers for next iteration
        \State $a\_ptrs, b\_ptrs \gets \text{UpdatePointers}(a\_ptrs, b\_ptrs, block\_k, split\_k)$
    \EndFor
    \State $\text{tl.atomic\_add}(c\_ptrs, acc)$ \Comment{Synchronization of partial sums across thread blocks}
    \end{algorithmic}
\end{algorithm}

\section{Performance}

We focused our experiments on M = 1 - 16, as this corresponds to a batch size of 1 - 16, a typical batch size range we might see in LLM inference. As our main contribution is a fused int4 dequantization kernel with a modified decomposition strategy, we explore the performance of the SplitK kernel vs the traditional DP kernel. The key difference is that in the DP decomposition, a single thread block is singly responsible for calculating the aggregate multiply-add accumulation, which produces an output tile in the C matrix. \ref{fig:data_parallel_drawing} We tested on both NVIDIA A100 and NVIDIA H100.

\subsection{SplitK vs Data Parallel Results}

We performed our performance analysis on a variety of different GPUs, with M = 1 - 16 and varying n = k. We conducted our tests on the NVIDIA A100 80GB SXM, A100 40GB PCIe and H100 PCIe to get a more comprehensive set of results. For both batch sizes and across all GPUs, SplitK performed much better than the DP decomposition.

\subsection{M=1}

\begin{figure}[H]
    \centering
    \begin{minipage}{0.48\textwidth}
        \centering
        \includegraphics[width=\textwidth]{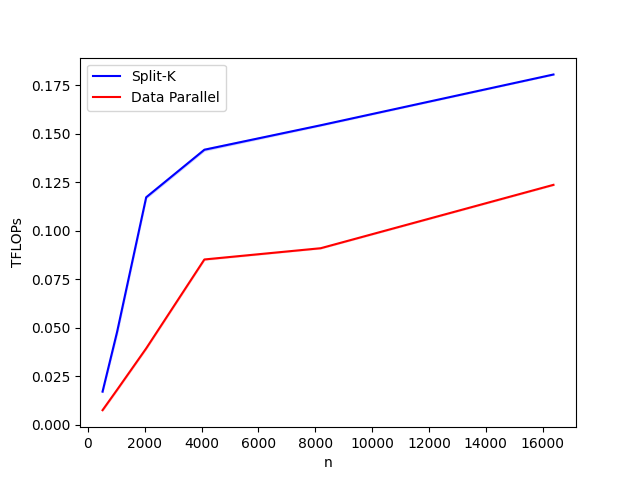}
        \caption{SplitK vs Data Parallel TFLOPS A100 40GB}
        \label{fig:split_k_vs_dp}
    \end{minipage}
    \hfill
    \begin{minipage}{0.48\textwidth}
        \centering
        \begin{tabular}{|c|c|c|c|}
            \hline
            N     & K     & SplitK [TFLOPS] & Data Parallel [TFLOPS]    \\ \hline
            512   & 512   & 0.01              & 0.07                    \\ \hline
            1024  & 1024  & 0.04              & 0.04                    \\ \hline
            2048  & 2048  & 0.11              & 0.08                    \\ \hline
            4096  & 4096  & 0.14              & 0.09                    \\ \hline
            8192  & 8192  & 0.15              & 0.09                   \\ \hline
            16384 & 16384 & 0.18              & 0.12                   \\ \hline
        \end{tabular}
        \captionof{table}{SplitK vs Data Parallel TFLOPS A100 40GB}
        \label{tab:split_k_vs_dp}
    \end{minipage}
\end{figure}

\begin{figure}[H]
    \centering
    \begin{minipage}{0.48\textwidth}
        \centering
        \includegraphics[width=\textwidth]{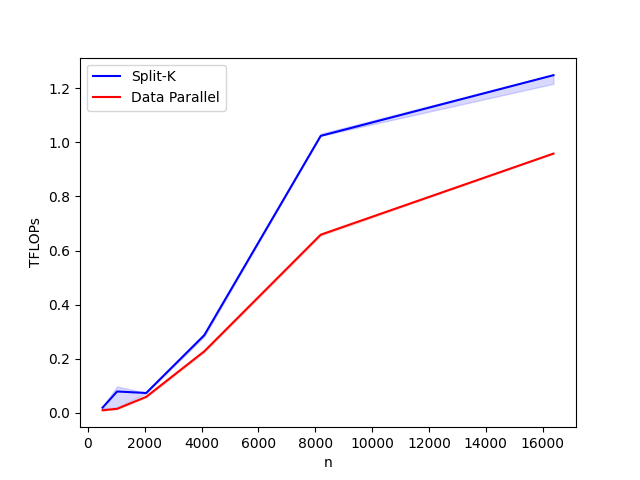}
        \caption{SplitK vs Data Parallel TFLOPS A100 80GB}
        \label{fig:split_k_vs_dp}
    \end{minipage}
    \hfill
    \begin{minipage}{0.48\textwidth}
        \centering
        \begin{tabular}{|c|c|c|c|}
            \hline
            N     & K     & SplitK [TFLOPS]   & Data Parallel [TFLOPS] \\ \hline
            512   & 512   & 0.02             & 0.01                    \\ \hline
            1024  & 1024  & 0.01             & 0.01                    \\ \hline
            2048  & 2048  & 0.06             & 0.04                    \\ \hline
            4096  & 4096  & 0.22             & 0.18                   \\ \hline
            8192  & 8192  & 1.03             & 0.66                   \\ \hline
            16384 & 16384 & 1.25             & 0.96                   \\ \hline
        \end{tabular}
        \captionof{table}{SplitK vs Data Parallel TFLOPS A100 80GB}
        \label{tab:split_k_vs_dp}
    \end{minipage}
\end{figure}

\begin{figure}[H]
    \centering
    \begin{minipage}{0.48\textwidth}
        \centering
        \includegraphics[width=\textwidth]{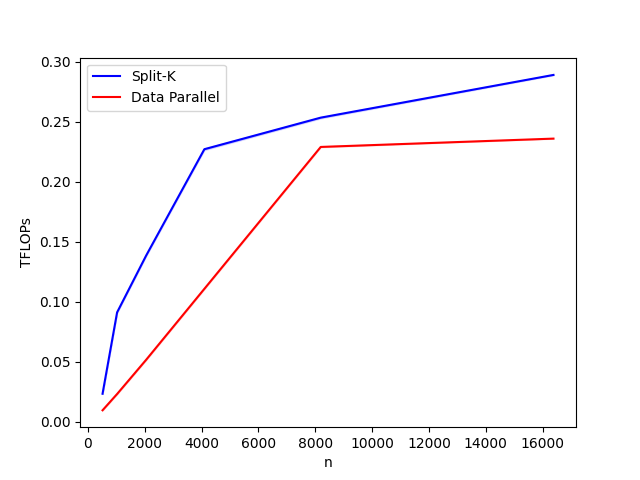}
        \caption{SplitK vs Data Parallel TFLOPS H100}
        \label{fig:split_k_vs_dp}
    \end{minipage}
    \hfill
    \begin{minipage}{0.48\textwidth}
        \centering
        \begin{tabular}{|c|c|c|c|}
            \hline
            N     & K     & SplitK [TFLOPS] & Data Parallel [TFLOPS] \\ \hline
            512   & 512   & 0.28            & 0.12                    \\ \hline
            1024  & 1024  & 0.77            & 0.28                    \\ \hline
            2048  & 2048  & 1.85            & 0.62                    \\ \hline
            4096  & 4096  & 2.25            & 1.36                    \\ \hline
            8192  & 8192  & 2.46            & 1.45                   \\ \hline
            16384 & 16384 & 2.87            & 1.98                   \\ \hline
        \end{tabular}
        \captionof{table}{SplitK vs Data Parallel TFLOPS H100}
        \label{tab:split_k_vs_dp}
    \end{minipage}
\end{figure}

\subsection{M=16}

\begin{figure}[H]
    \centering
    \begin{minipage}{0.48\textwidth}
        \centering
        \includegraphics[width=\textwidth]{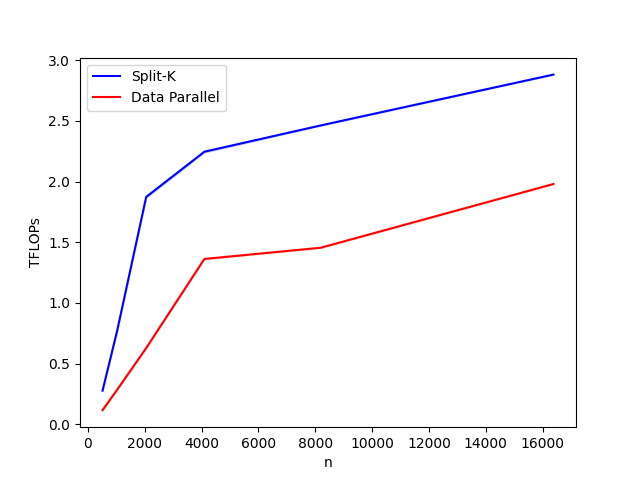}
        \caption{SplitK vs Data Parallel TFLOPS A100 40GB}
        \label{fig:split_k_vs_dp}
    \end{minipage}
    \hfill
    \begin{minipage}{0.48\textwidth}
        \centering
        \begin{tabular}{|c|c|c|c|}
            \hline
            N     & K     & SplitK [TFLOPS] & Data Parallel [TFLOPS] \\ \hline
            512   & 512   & 0.3              & 0.1                    \\ \hline
            1024  & 1024  & 0.8              & 0.3                    \\ \hline
            2048  & 2048  & 1.9              & 0.6                    \\ \hline
            4096  & 4096  & 2.2              & 1.4                    \\ \hline
            8192  & 8192  & 2.5              & 1.5                   \\ \hline
            16384 & 16384 & 2.9              & 2.0                   \\ \hline
        \end{tabular}
        \captionof{table}{SplitK vs Data Parallel TFLOPS A100 40GB}
        \label{tab:split_k_vs_dp}
    \end{minipage}
\end{figure}

\begin{figure}[H]
    \centering
    \begin{minipage}{0.48\textwidth}
        \centering
        \includegraphics[width=\textwidth]{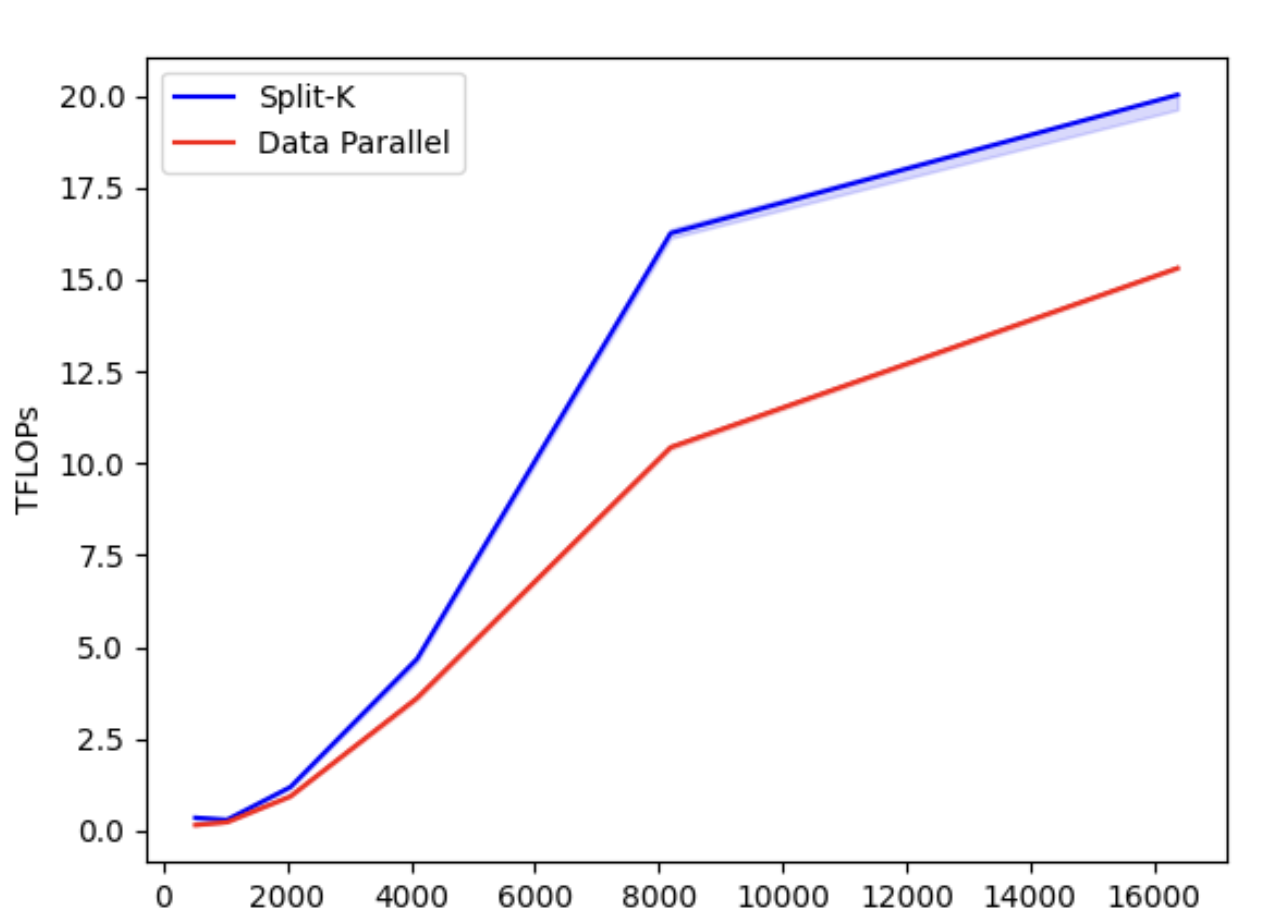}
        \caption{SplitK vs Data Parallel TFLOPS A100 80GB}
        \label{fig:split_k_vs_dp}
    \end{minipage}
    \hfill
    \begin{minipage}{0.48\textwidth}
        \centering
        \begin{tabular}{|c|c|c|c|}
            \hline
            N     & K     & SplitK [TFLOPS] & Data Parallel [TFLOPS]   \\ \hline
            512   & 512   & 0.3              & 0.1                     \\ \hline
            1024  & 1024  & 0.3              & 0.2                     \\ \hline
            2048  & 2048  & 1.1              & 0.9                     \\ \hline
            4096  & 4096  & 4.5              & 3.5                     \\ \hline
            8192  & 8192  & 16.3             & 10.4                    \\ \hline
            16384 & 16384 & 20.0             & 15.3                    \\ \hline
        \end{tabular}
        \captionof{table}{SplitK vs Data Parallel TFLOPS A100 80GB}
        \label{tab:split_k_vs_dp}
    \end{minipage}
\end{figure}

\begin{figure}[H]
    \centering
    \begin{minipage}{0.48\textwidth}
        \centering
        \includegraphics[width=\textwidth]{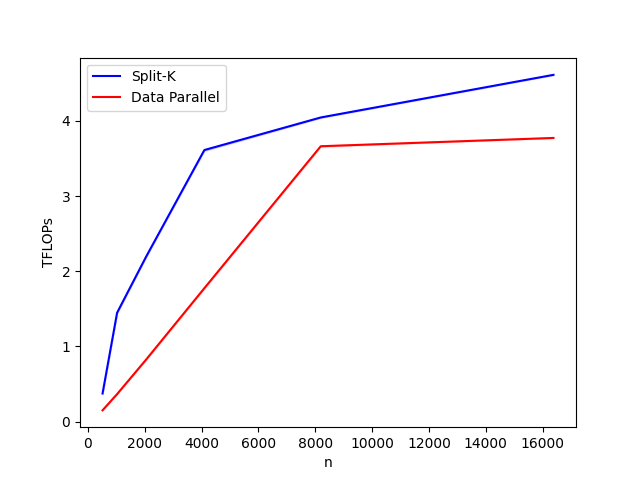}
        \caption{SplitK vs Data Parallel TFLOPS H100}
        \label{fig:split_k_vs_dp}
    \end{minipage}
    \hfill
    \begin{minipage}{0.48\textwidth}
        \centering
        \begin{tabular}{|c|c|c|c|}
            \hline
            N     & K     & SplitK [TFLOPS] & Data Parallel [TFLOPS]  \\ \hline
            512   & 512   & 0.4              & 0.2                    \\ \hline
            1024  & 1024  & 1.4              & 0.2                    \\ \hline
            2048  & 2048  & 2.2              & 0.9                    \\ \hline
            4096  & 4096  & 3.6              & 1.7                    \\ \hline
            8192  & 8192  & 4.1              & 3.7                    \\ \hline
            16384 & 16384 & 4.6              & 3.8                   \\ \hline
        \end{tabular}
        \captionof{table}{SplitK vs Data Parallel TFLOPS H100}
        \label{tab:split_k_vs_dp}
    \end{minipage}
\end{figure}

The average speedup for the SplitK strategy is 1.24x, 1.14x and 0.64x on the H100, A100 40GB and A100 80GB respectively. We also tested a variety of SplitK settings and found varying optimal tile splitting factors. We found split\_k = 4 and and split\_k = 8 produced the best results on the A100 80GB and H100 respectively. For these tests, we fixed the tile sizes, number of warps, and number of stages to isolate the impact of SplitK.

\begin{figure}[H]
    \centering
    \begin{minipage}[b]{0.48\textwidth}
        \includegraphics[width=\textwidth]{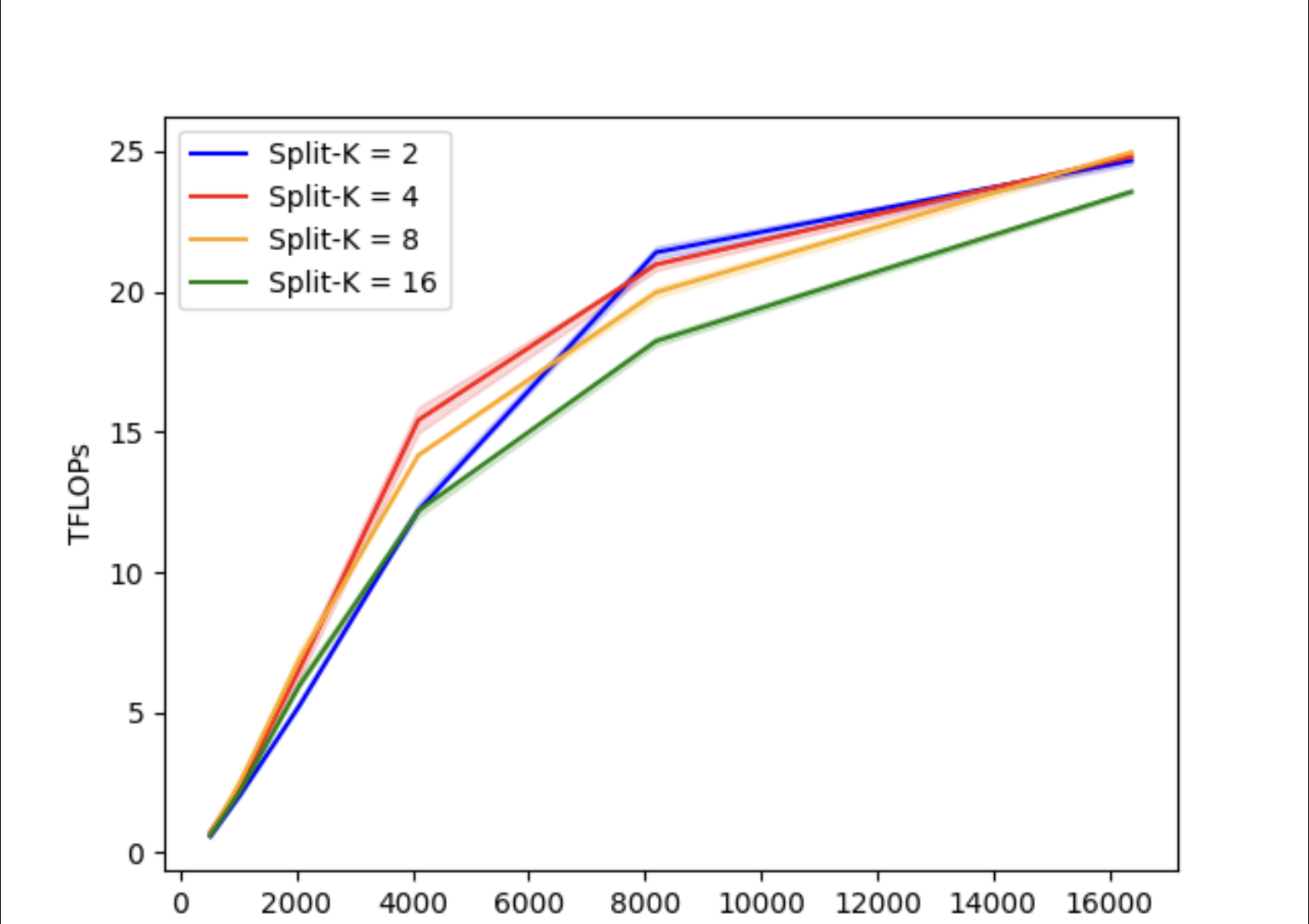}
        \caption{SplitK Comparison of splitting factor, A100}
        \label{fig:split_k}
    \end{minipage}
    \hfill
    \begin{minipage}[b]{0.48\textwidth}
        \includegraphics[width=\textwidth]{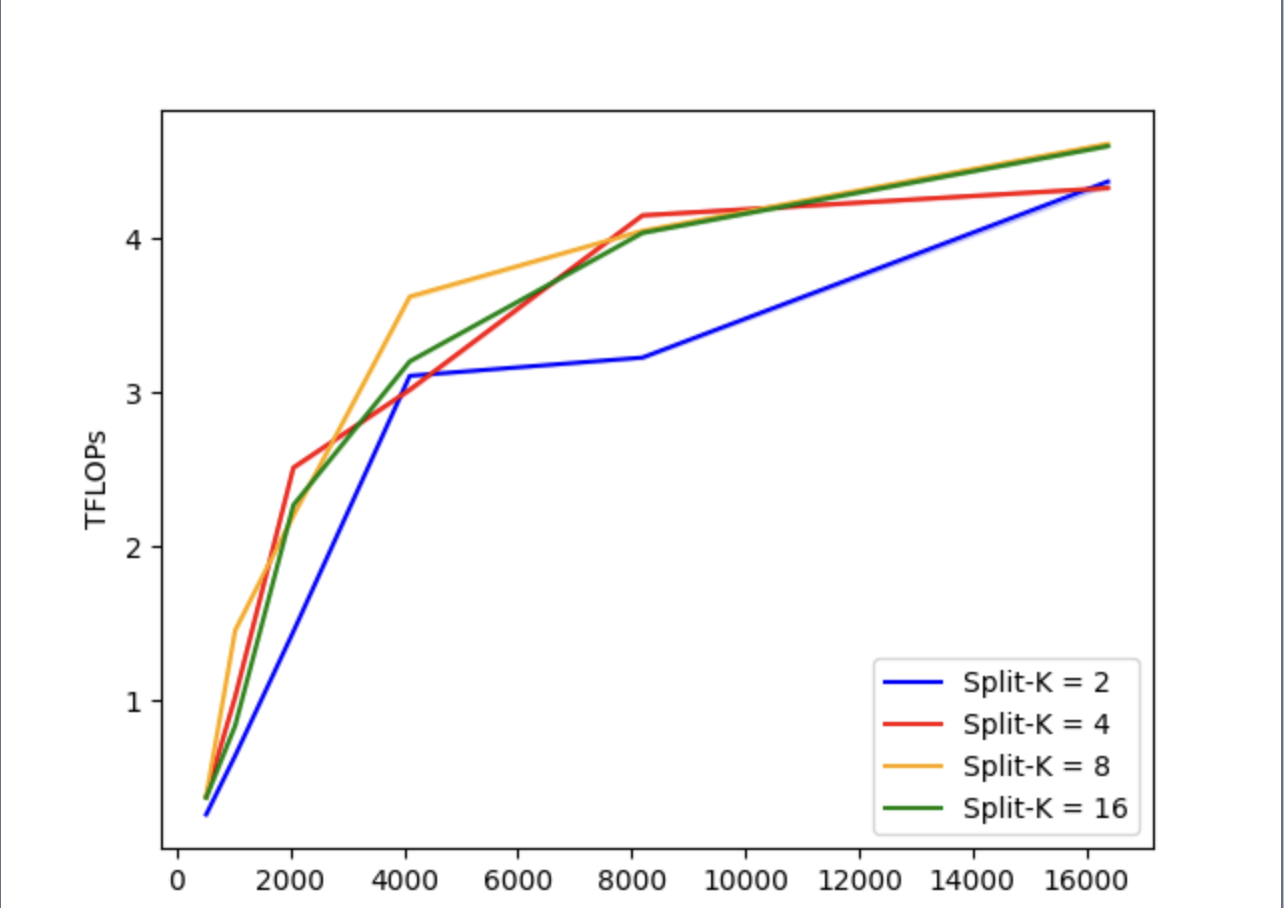}
        \caption{SplitK Comparison of splitting factor,  H100}
        \label{fig:data_parallel}
    \end{minipage}
\end{figure}

\subsection{SplitK vs Data Parallel NVIDIA Nsight Compute}

Statistics gathered from NVIDIA Nsight Compute \cite{noauthor_nsight_nodate}, shown below, suggest that the SplitK kernel performs better than the DP Kernel mainly due to increased global memory throughput. These metrics were collected for a single test case where m=16, n= 4096 and k=4096. 

\begin{table}[H]
\centering
\captionsetup{position=bottom}
\begin{tabular}{|l|c|c|}
\hline
\textbf{Metrics}              & \textbf{SplitK} & \textbf{Data Parallel}    \\ \hline
Latency                       & 27.90us              & 52.93us              \\ \hline
Global Memory Throughput      & 313 GB/s             & 161 GB/s             \\ \hline
Grid Size                     & 512                  & 128                  \\ \hline
Registers                     & 92                   & 150                  \\ \hline
Shared Memory Usage           & 102.40KB             & 167.94KB             \\ \hline
Block Limit (Registers)       & 5                    & 3                    \\ \hline
Block Limit (SMEM)            & 5                    & 2                    \\ \hline
Achieved Occupancy            & 27.75                & 7.55                \\ \hline
SM Utilization                & 43.05\%              & 20.75\%            \\ \hline
\end{tabular}
\caption{Nsight Compute Metrics}
\label{tab:SplitK vs Data Parallel NCU Statistics}
\end{table}

The SplitK kernel launches 4x more thread blocks than the DP Kernel. Each thread block is responsible for calculating a partial result of an output tile as opposed to a complete tile. The thread blocks in the SplitK kernel then, notably have less work to do compared to the DP Kernel, which leads to better load balancing across SMs. This intuition can be directly confirmed by the 2x improvement in SM utilization.

\begin{figure}[H]
    \centering
    \begin{minipage}[b]{0.48\textwidth}
        \includegraphics[width=\textwidth]{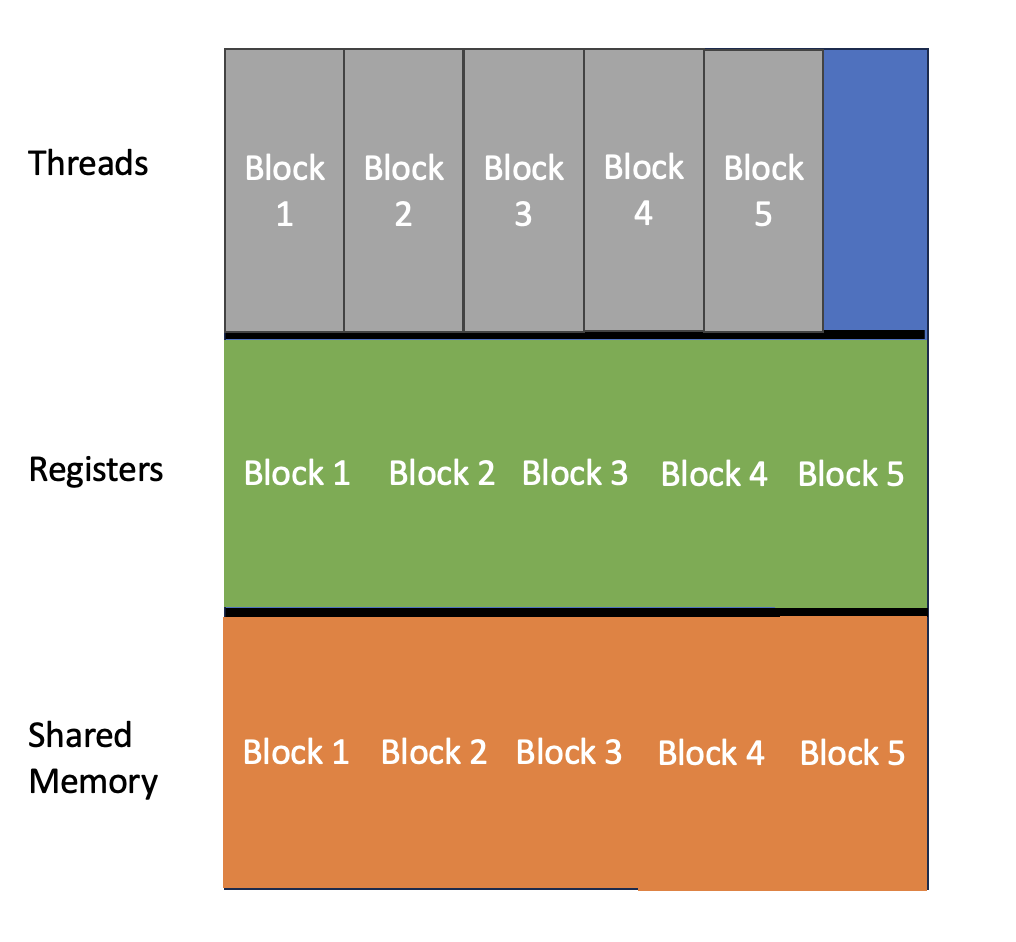}
        \caption{SplitK A100 SM Resource Usage}
        \label{fig:split_k_sm}
    \end{minipage}
    \hfill
    \begin{minipage}[b]{0.48\textwidth}
        \includegraphics[width=\textwidth]{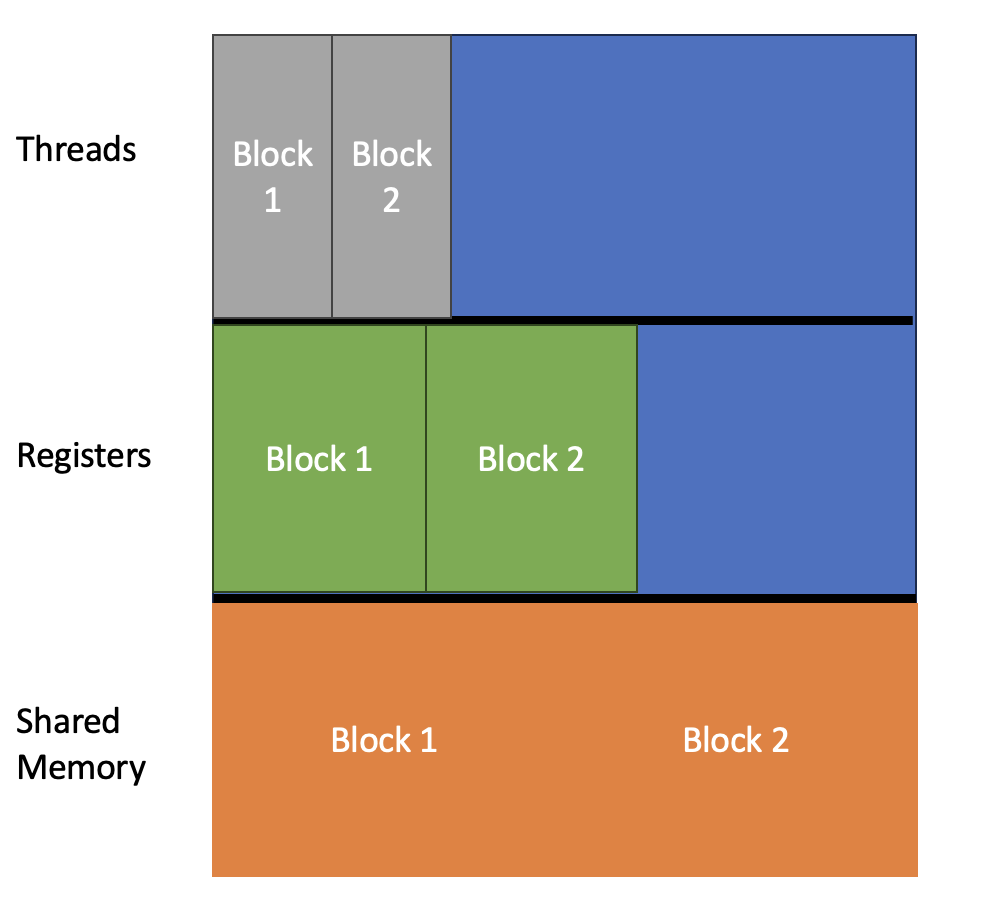}
        \caption{Data Parallel A100 SM Resource Usage}
        \label{fig:data_parallel_sm}
    \end{minipage}
\end{figure}

We note that the decreased register and shared memory usage in the SplitK kernel results in a nearly 4x improvement in occupancy.  The DP kernel, by contrast, is shared memory limited. This means that the shared memory resource requirements of each thread block limits the amount of active thread blocks on an SM. In contrast, the SplitK kernel has thread block limit of 5, a 2.5x improvement due to reduced shared memory and register use. This difference is depicted in \ref{fig:split_k_sm} and \ref{fig:data_parallel_sm}. The increase in occupancy and global memory throughput are both important metrics in understanding the latency improvement in the SplitK kernel.  The increase in both metrics are roughly proportional to the improvement in latency.  We may also use the Warp Scheduler Statistics to further understand the latency hiding characteristics of the SplitK kernel.

\begin{table}[H]
\centering
\captionsetup{position=bottom}
\begin{tabular}{|l|c|c|}
\hline
\textbf{Metrics}              & \textbf{SplitK} & \textbf{Data Parallel}    \\ \hline
Active Warps& 4.45            & 1.21                      \\ \hline
Eligible Warps                & 0.67            & 0.20                      \\ \hline
Issued Warps                  & 0.43            & 0.19                      \\ \hline
Issued IPC Active             & 1.72            & 0.75                      \\ \hline
\end{tabular}
\caption{Scheduler Statistics}
\label{tab:SplitK vs Data Parallel NCU Statistics}
\end{table}

The SplitK kernel has 2x more warps in the issued slot. This illustrates the algorithms superior ability to hide memory latency. The increase in issued slot utilization is able to increase our compute pipeline utilization. This can be seen by the 1.3x increase in issued instructions per clock cycle (IPC).  

\subsection{H100 vs A100}

We note the following performance specifications of the A100 \cite{noauthor_nvidia_nodate-1} and H100 \cite{noauthor_nvidia_nodate}, as our results indicated a varied set of optimal hyper parameters across tests. These key specifications can be used to help understand those differences.

\begin{table}[H]
\centering
\captionsetup{position=bottom}
\begin{tabular}{|l|c|c|c|}
\hline
\textbf{Feature}        & \textbf{NVIDIA H100 80GB PCIe} & \textbf{NVIDIA A100 80GB SXM} & \textbf{NVIDIA A100 40GB PCIe}       \\ \hline
Architecture            & Hopper                         & Ampere                        &  Ampere                             \\ \hline
SMs                     & 132                            & 108                           &  108                                \\ \hline
FP16 Tensor Core        & 1513 TFLOPS                    & 312 TFLOPS                    &  312 TFLOPS                         \\ \hline
Memory                  & 80GB HBM3                      & 80 GB HBM2                    &  40GB HBM2                          \\ \hline
Memory Bandwidth        & 2.0 TB/s                       & 2.0 TB/s                      &  1.5TB/s                            \\ \hline
L2 Cache                & 50MB                           & 40MB                          &  40MB                               \\ \hline
L1 Cache/SM             & 256KB                          & 192KB                         &  192KB                              \\ \hline
\end{tabular}
\caption{Comparison of NVIDIA H100 and A100 GPUs}
\label{tab:gpu_comparison}
\end{table}

Notably, the smaller form factor A100 sees an increased speedup when using SplitK, compared to the larger form factor (1.14x vs 0.64x).

This is intuitively explained by the fact that on the smaller A100, the memory bandwidth speed is 31\% slower, and thus for small problem sizes, the kernels are comparatively more memory bound than when run on the larger (faster memory bandwidth) A100s. The finer decomposition and larger grid of the SplitK algorithm thus provides proportionately greater benefit (1.14x vs 0.64x) relative to the A100 80GB with higher memory bandwidth.

\section{Conclusion}
We present an optimized Triton kernel implementation for quantized matrix-matrix multiplications in inference workloads, where the problem is memory bound. Our implementation is a fused kernel that performs dequantization and GEMM via SplitK atomic reduction. We benchmark m, n and k values that are relevant for llama-style inference inputs and show an average of 65\% speed improvement on A100, and 124\% speed improvement on H100 (with a peak of 295\%) compared to traditional blocked data parallelization strategies.  We provide kernel level analysis to explain the speedups as being driven by a finer grained work decomposition via SplitK algorithm.  This results in greater SM occupancy, and thus improved global memory throughput via latency hiding, as well as reduced wave quantization inefficiency. A potential future direction would be to explore the natural successor to SplitK, StreamK. The StreamK \cite{osama_stream-k_2023} decomposition may potentially enable even finer grained optimal work decomposition resulting in additional performance improvements for GEMM workloads. 

\printbibliography

\end{document}